\documentclass[fleqn,twoside]{article} 
\usepackage{epsf,multicol,ifthen}
\usepackage{ujp}
\usepackage[cp1251]{inputenc}
\usepackage[english]{babel}
\usepackage{amstext}
\nazva{The CDF-II Tau Physics Program\\
Triggers, $\tau$ ID and Preliminary Results}%

\udk{preprint }

\nazvacol{The CDF-II Tau Physics Program}%

\avtor{S.~Baroiant$\footnotesize^1$, M.~Chertok$\footnotesize^1$,
M.~Goncharov$\footnotesize^2$ T.~Kamon$\footnotesize^2$,
V.~Khotilovich$\footnotesize^2$, R.~Lander$\footnotesize^1$,\,\,\,
T.~Ogawa$\footnotesize^3$,\,\,\,\,
\underline{C.~Pagliarone}$\footnotesize^4$,\,\,\,\,
F.~Ratnikov$\footnotesize^5$,\,\,\, A.~Safonov$\footnotesize^1$,\\
A.~Savoy-Navarro$\footnotesize^6$,\,\,\,
J.R.~Smith$\footnotesize^1$,\,\,\, E.~Vataga$\footnotesize^4$\\
\vskip5pt
(For the CDF-II Collaboration)}%

\avtorcol{Carmine Pagliarone }%

\inst{University of California, Davis}%
 \adr{(Davis, California, 95616 - USA), }
\insti{Texas A\&M University}%
 \adri{(College Station, Texas 79409 - USA), }%
\instii{Waseda University}%
 \adrii{(169-8555 Tokyo - Japan) }%
\instiii{INFN Pisa}%
 \adriii{(via F.Buonarroti, 2 - 56100 PISA - Italy), }%
\instiiii{Rutgers University}%
 \adriiii{(Piscataway, New Jersey 08854 - USA), }%
\instiiiii{LPNHE, University of Paris 6-7 and CNRS-IN2P3 }%
 \adriiiii{(Paris - France), }%

\begin{document}           
\setcounter{page}{1}%
\maketitl                 
\begin{multicols}{2}

\anot{%
{\bf\underline{Abstract}}\\

\noindent The study of processes containing $\tau$ leptons in the
final state will play an important role at Tevatron Run II. Such
final states will be relevant both for electroweak studies and
measurements as well as in searches for physics beyond the
Standard Model. The present paper discuss the physics
opportunities and challenges related to the implementation of new
set of triggers able to select events containing tau candidates in
the final state. We illustrate, in particular, the physics
capabilities for a variety of new physics scenarios such as
supersymmetry (SUSY), SUSY with $\mathcal{R_{P}}$-parity
violation, with Bilinear parity violation or models with the
violation of lepton flavor. Finally, we present the first Run II
results
obtained using some of the described tau triggers.}%

\section{Introduction}

Since the discovery of the $\tau$ lepton in
1975~\cite{taudiscovery} the knowledge about its properties and
interactions has improved drastically. At present most of the tau
physics is shifting towards high precision measurements involving
the determination of the $\tau$ static properties, the
investigation of the lepton universality and tests of the Lorentz
structure of $\tau$ decays. The tau physics program at Tevatron
collider is predominantly focused on accessing processes
containing taus~\cite{Abel}.

\noindent The study of events with $\tau$ leptons in the final
state will play an important role in Tevatron Run II, both for
electroweak studies and measurements both for searches for physics
beyond the Standard Model. In order to be able to select such
events, specific dedicated tau trigger systems have been designed
and implemented in the Collider Detector at Fermilab (CDF-II)
experiment. This article discusses the physics opportunities and
challenges related to the implementation of this set of $\tau$
triggers. The paper is organized as follows first we review the
major CDF-II upgrades; then we discuss which are the typical
signatures for a tau lepton at Tevatron; next follows an
illustration of the tau Trigger architecture and the variety of
physics topics addressable by the Lepton Plus Track (LPT) Trigger.
Finally we describe some preliminary result based on data samples
collected by using the LPT trigger.

\section{The CDF-II Detector}

CDF-II is a $5000$ ton multi-purpose particle physics
experiment~\cite{CDF-1} dedicated to the study of
proton-antiproton collisions at the Fermilab Tevatron collider. It
was designed, built and operated by a team of physicists,
technicians and engineers that by now spans over $44$ institutions
and includes, approximately, more than $500$  members. The history
of the experiment goes back over $20$ years. The CDF detector has
been recently upgraded~\cite{CDF-upgrades} in order to be able to
operate at the high radiation and high crossing rate of the Run II
Tevatron environment. In addition, there have been several
upgrades to improve


\vspace{0.3 cm} \noindent \vskip17pt
\hspace{-0.6cm}\epsfxsize=\columnwidth\epsffile{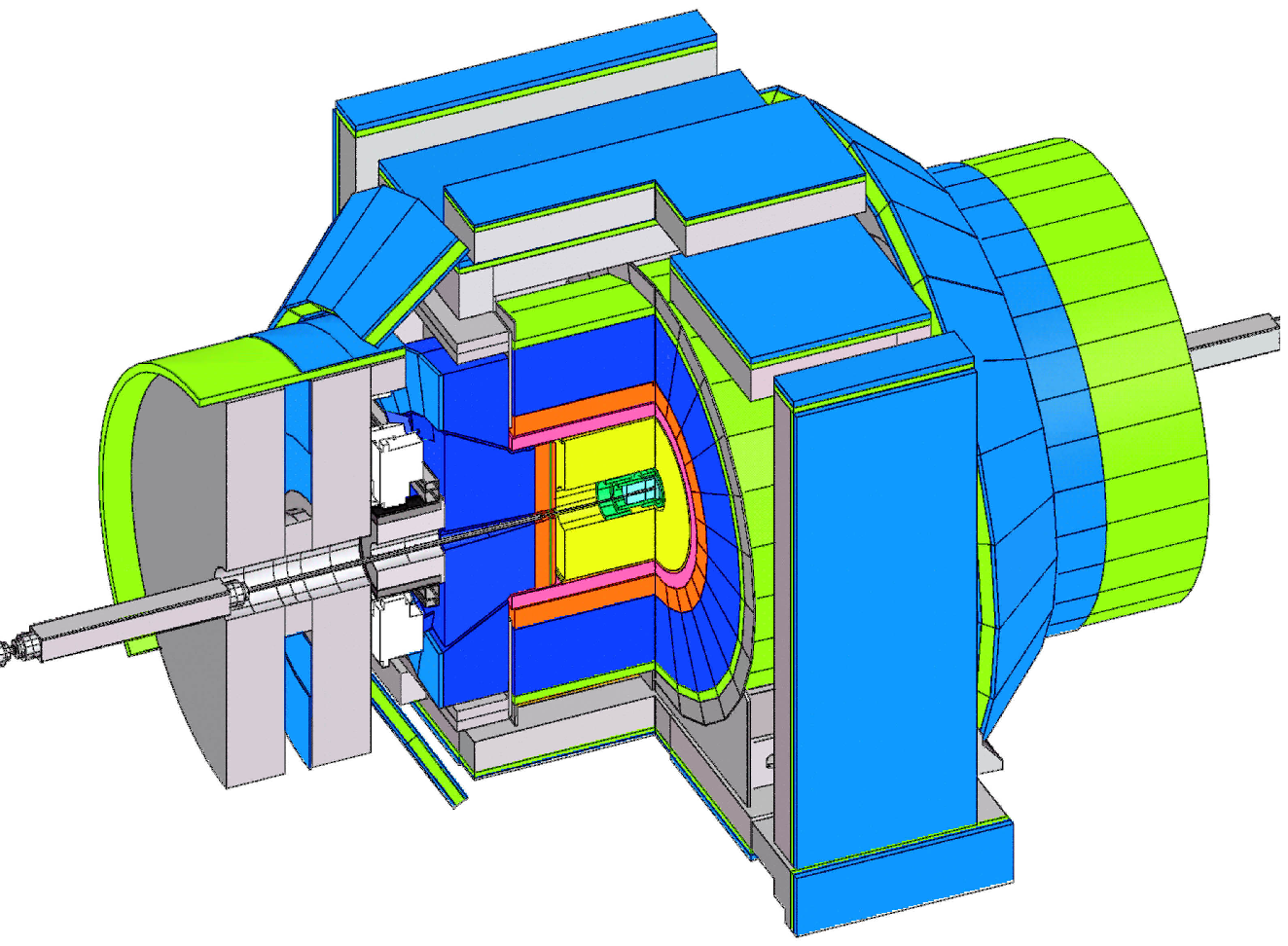}
\vskip3pt \noindent Figure 1:\,\, {\it An overview of the Collider
Detector at Fermilab (CDF) in its Run II configuration (CDF-II).}
\noindent{\footnotesize } \vskip40pt

\noindent  the sensitivity of the detector to specific physics
tasks such as heavy flavor physics, Higgs boson searches and many
others. Figure~1 shows an isometric cutaway view of the final
configuration of the CDF-II detector. The central tracking volume
of the CDF experiment has been replaced entirely with new
detectors (see Figure~2), the central calorimeters has not been
changed. These upgrades can be summarized as follows:

\noindent A new {\bf Silicon System}  done of $3$ different
tracking detector subsystems:

\noindent {\bf Layer00} -- a layer of silicon detectors installed
directly on the beam pipe to increase impact parameter resolution.

\noindent {\bf Silicon Vertex Detector (SVX II)} -- to meet new
physics goals, a central vertexing portion of the detector called
SVX II was designed. It consists of double-sided silicon sensors
with a combination of both 90-degree and small-angle stereo
layers. The SVX II is nearly twice as long as the original SVX and
SVX$^{\prime}$ ($96$ $cm$ instead of $51$ $cm$), which were
constrained to fit within a previous gas-based track detector
(CTC) used to locate the position of interactions along the beam
line. SVX II has 5 layers instead of $4$ of the previous silicon
detector and it is able to give 3-dimensional information on the
tracks at trigger level.

\noindent {\bf Intermediate Silicon Layer (ISL)}
 is a large radius ($R= 29$ $cm$)
silicon tracker with a total active area of $\sim 3.5\,\,m^{2}$.
It is composed of $296$ basic units, called ladders, made of three
silicon sensors bonded together in \,\,order \,\,to\,\, form\,\,
one\, electric\,\,unit. \,\,The \,\,ISL \,is\, \,located

\noindent \vskip20pt
\hspace{-0.5cm}\epsfxsize=\columnwidth\epsffile{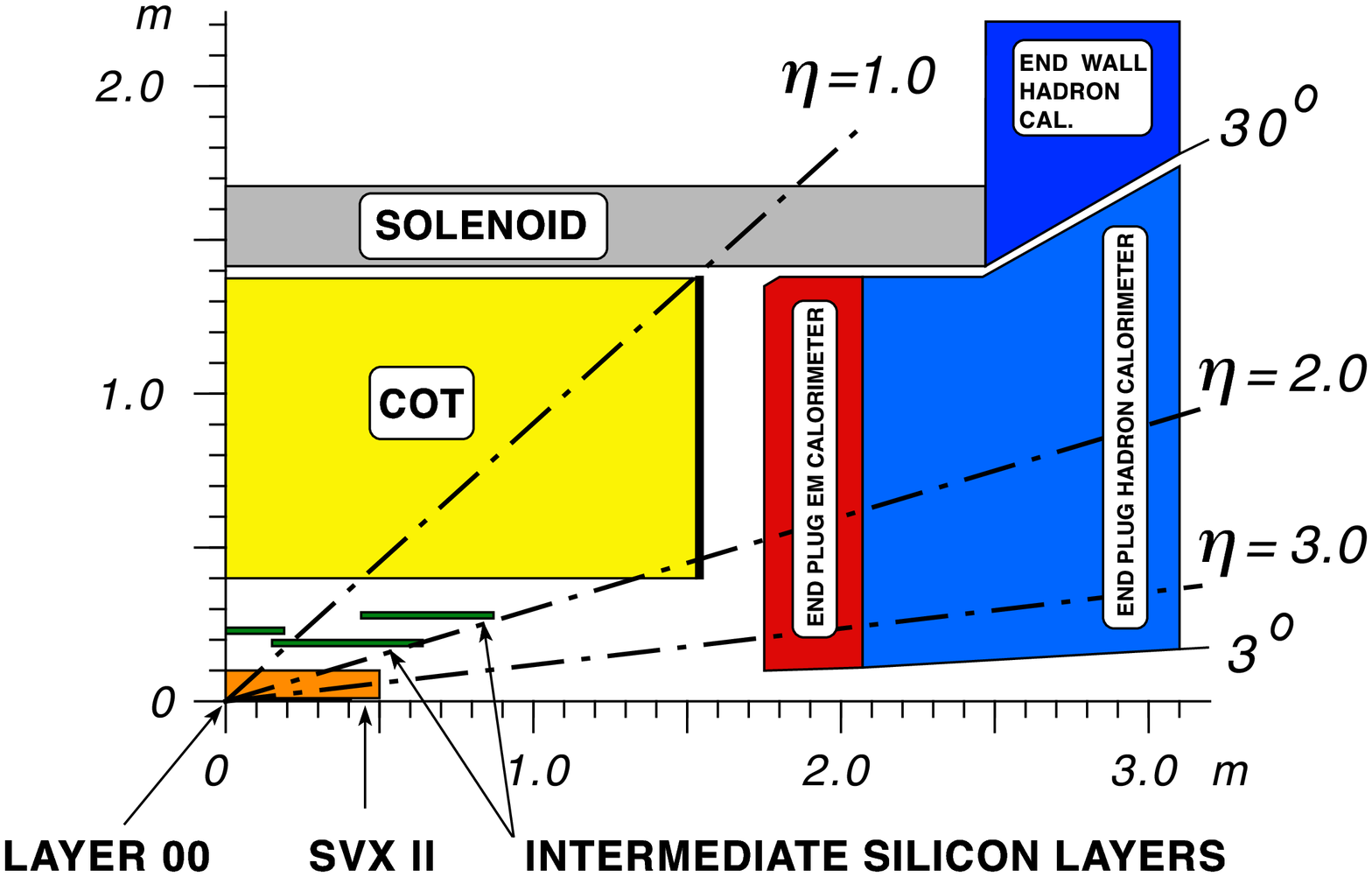}
 \label{d2}
\vskip9pt

\noindent Figure 2: \,\,{\it A cutaway view of one quadrant of the
inner portion of the CDF-II detector showing the tracking region
surrounded by the solenoid and endcap calorimeters.}
\noindent{\footnotesize } \vskip12pt

\noindent between the Silicon Vertex Detector and the Central
Outer Chamber. Being at a distance of $\sim 23$ $cm$ in the
central part, from the beam-line, it increase the pseudorapidity
reach of tracking system up to $|\eta|\,<\,2$.

\noindent {\bf Central Outer Tracker (COT)}\\ The COT is the new
CDF central tracking chamber. It is an open cell drift chamber
able to operate at a beam crossing time of 132~$ns$ with a maximum
drift time of $\sim\!100$~$ns$. The COT consists of 96 layers
arranged in four axial and four stereo superlayers. It also
provides d$E$/d$x$ information for particle identification.\\

\noindent {\bf Time-of-Flight Detector (TOF)}\\ New scintillator
based Time-of-Flight detector has been added using a small space
available between COT and solenoid. With its expected $100$ $ps$
time-of-flight resolution, the TOF system will enhance the
capability to tag charged kaons in the $P_{\rm T}$ range from
$\sim 0.6$ to few GeV$/c$ as requested from the $B$ physics
program.\\

\noindent {\bf Plug Calorimeter}\\ A new scintillating tile plug
calorimeter has been realized in order to have a good electron
identification up to $|\eta|=2$.\\

\noindent {\bf Muon system} has also been upgraded: the coverage
in the central region has been almost doubled.\\

\noindent A new {\bf Data Acquisition System (DAQ)} has been

\noindent \epsfxsize=\columnwidth\epsffile{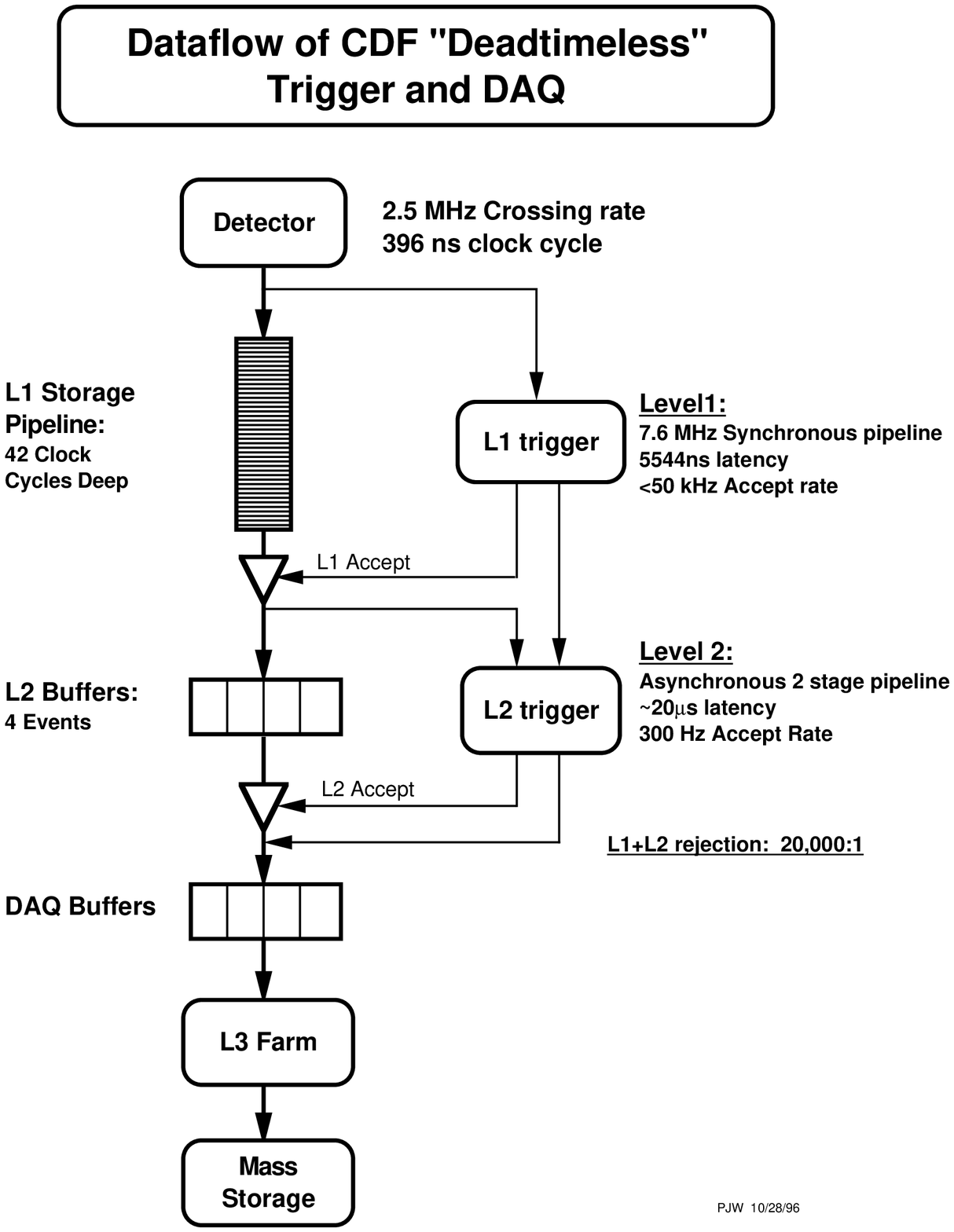}
\vskip20pt \noindent Figure 3: \,\,{\it Data flow of CDF-II
Trigger and Data Acquisition architecture.}
\noindent{\footnotesize }%
\vskip30pt
\noindent adapted to short bunch spacing of 132 ns. It is capable
to record data with event size of the order of 250 KB and
permanent logging of 20 MB/s.\\

\section{The CDF-II Trigger System}

The CDF-II trigger system has been completely rebuilt for Run II
needs. The present architecture as well as the previous one has a
three-level structure, that is used to reduce the 7.6 MHz nominal
crossing rate to 50 Hz maximum written to tape. Using a pipelined
and buffered system, the trigger is designed to be "deadtimeless".
The structure of the trigger is flexible and programmable in order
to respond to changing beam conditions or physics goals. Level 1
(L1) uses custom designed hardware to find physics objects with a
subset of the detector information and makes a decision based on
simple counting of these objects: for example, $\mu$ or $e$ with
$P_T>$ 6~GeV$/c$. The most significant upgrade in the L1 trigger
for Run II is the addition of track finding: a new fast trigger
processor, called eXtremely Fast Trigger (XFT)\cite{XFT} find
tracks in the central drift chamber (COT) in the $r$-$\phi$ plane
with a resolution of $\delta P_{\rm T}/P_{\rm T}
> 1.6$ \% and $\delta\phi=4\;\mu$rad in $2.7\;\mu$sec after a
collision. The XFT also matches tracks to muon stubs or
calorimeter towers, and the Level 1 decision is generated in
$4\;\mu$s with accept rate about 50 kHz. Level 2 (L2) operates
after detector readout and has an accept rate in the range 200 to
300 Hz. All the information used in the Level 1 decision is
available to the Level 2 system, but with higher precision. Track
collection available at Level 2 is 2-dimensional and includes
tracks with $P_{\rm T}>$ 1.5~GeV$/c$. In addition, jet
reconstruction is made by a Level 2 cluster finder. The Level 3
(L3) trigger benefits of the full detector resolution including
3-D track reconstruction and tight matching of tracks to
calorimeter and muon-system information. Its algorithm structure
is based on the CDF offline analysis software and is used to
reconstruct events in a processor farm with a final accept rate of
50 to 70 Hz.

\section{Tau Identification}

Tau leptons decay predominantly into charged and neutral hadrons
as showed in Table~\ref{tab:modes}. Hadronic decays happen with a
total branching ratio that is $\sim$ 64\%. There is, always, at
least, one $\nu$ in the final state. This kind of $\tau$ decays
have the distinct signature of a narrow isolated jet with low
charged track multiplicity
and low visible mass ($M_{\rm visible}\,<\,M_{\tau}$). Taus decay
also leptonically with a branching ratio that is:
$\mathcal{BR}(\tau \rightarrow \nu \nu e) \sim$ $18$\% for the
electron channel and $\mathcal{BR}(\tau \rightarrow \nu \nu
\mu)\sim$ $17$\% for the muon channel.
The presence of at least one neutrino, in all possible tau decay,
implies that energy cannot be measured directly. However, the
direction of a tau can be obtained from the observed decay
products as the energy of such decay products is large compared to
the mass. There is also another practical correlation between the
tau visible energy and the cone size of the tracks. The bigger is
the energy of the tau lepton, the smaller will be the cone size
defined by these escaping tracks. As a side effect, taus will
appear, in the rest frame of the CDF-II detector, as narrow
isolated jets, with low track multiplicity.
Then taus coming from $Z^{0}$ or $W$ decay show up to be
ultra-relativistic and to have mainly one or three prong's track
multiplicity. Then to enhance, at trigger level, the separation of
tau-jet signal from the generic $W\,+\,$jets background, an
isolation cone cut have to be applied.
The basis of our triggers is, therefore, a $\tau$-cone


\begin{table*}
\begin{center}
\begin{tabular}{l l l}
\hline
 $\tau\rightarrow e\nu\nu_\tau$          & $\approx 18\%  $  &
$\mathcal{BR}$, univ, Michel \\
 $\tau\rightarrow \mu\nu\nu_\tau$        & $\approx 17\%  $  &
$\mathcal{BR}$, univ, Michel \\
 $\tau\rightarrow \pi\nu , \, K\nu_\tau$ & $\approx 12\%  $  &
$\mathcal{BR}$, univ  \\
 $\tau\rightarrow \pi\pi\nu_\tau$        & $\approx 25\%  $  &
$\mathcal{BR}$, $\rho$, $\rho^\prime$, CVC, $\Pi$ \\
 $\tau\rightarrow K\pi\nu_\tau$          & $\approx 1.4\% $  &
$\mathcal{BR}$, $K^*$, $K^{*\prime}$ \\
 $\tau\rightarrow 3\pi\nu_\tau$          & $\approx 18\%  $  &
$\mathcal{BR}$, $a_1$, $a_1^\prime$\\
 $\tau\rightarrow K\pi\pi\nu_\tau$       & $\approx 0.8\% $  &
$\mathcal{BR}$, $K_1$, $K_{1b}$, $W$-$Z^{0}$ \\
 $\tau\rightarrow 4\pi\nu_\tau$          & $\approx 5\%   $  &
$\mathcal{BR}$, $\rho^\prime$, CVC  \\
  $\tau\rightarrow \mbox{rare}$     & $\approx 2\%   $  &
$5\pi$, $6\pi$, $\eta\pi\pi$, ... \\
  $\tau\to \eta\pi\nu, b_1\nu_\tau$       & $\ll 1\%     $  &
2$^{nd}$-class currents \\
  $\tau\rightarrow \mbox{forbidden}$ & $\ll 1\%     $  &
neutrinoless decays \\ \hline
\end{tabular}
\vskip5pt \caption{\label{tab:modes} Main decay modes of the
$\tau$ lepton~\cite{taucleo}.}
\end{center}
\end{table*}
\noindent algorithm for the reconstruction of hadronic $\tau$
decays. To build the $\tau$-cone object, we start with a narrow
calorimeter cluster above a suitable $E_{\rm T}$ threshold, which
determines the direction for searching the seed track with
momentum above certain $P_{\rm T}$.

\noindent The region within an angle $\Theta_{\rm Sig}$ from the
seed track direction is used to define a cone of tracks to be
associated with the candidate $\tau$ lepton. The region between
$\Theta_{\rm Sig}$ and $\Theta_{\rm Iso}$ defines the isolation
cone, and we require that no tracks with $P_{\rm T}$ higher than a
fixed low threshold have to be found in the isolation cone. At
trigger level, $\Theta_{\rm Sig}$ and $\Theta_{\rm Iso}$ have
values of $10^{\circ}$ ($0.1745$ rad ), and $30^{\circ}$ ($0.5235$
rad): $N^{10^{0}-30^{0}}_{track}=\,0$. At L2, these angles are
available in the $r$-$\phi$ plane, while at L3 it is possible to
use 3-dimensional angles. In offline tau reconstruction we allow
the $\tau$ cone to shrink with the increase of the visible energy
to take into account the Lorentz non invariance of fixed cone
clustering algorithm.

\section{The CDF-II Tau Triggers}

 \noindent The CDF-II $\tau$ Triggers are a set of Triggers
integrated into all 3 levels of the general CDF-II Trigger system.
At the present CDF-II has five different $\tau$ triggers
operating:
\begin{itemize}
\item Central Muon Plus Track;
\item CMX Muon Plus Track;
\item Central Electron Plus Track;
\item Di-Tau Trigger;
\item Tau + $\not\!\!\!E_{\rm T}$;
\end{itemize}
These Triggers were installed in the CDF-II trigger tables in
January 2002. Naturally, the design of these triggers has evolved
in time. At the present they are all working properly collecting
data in stable, non pre-scaled way. Below we will describe in more
details the basic characteristics of each of the LPT triggers.

\section{The Lepton Plus Track Triggers}

The Lepton Plus Track Trigger is a class of low momentum dilepton
triggers able to select events containing charged leptons,
including $\tau$'s, in the final state~\cite{cdf6325}.

\noindent As taus in $\sim\,35$ \% of cases promptly decay into
leptons and the rest of times in hadrons, then dilepton events,
where both leptons are $\tau$'s, can be identified by accessing
both purely leptonic di-$\tau$ decays: $\,\tau_{e} \,\tau_{e}\,$,
$\,\tau_{e} \, \tau_{\mu}\,$ or mixed leptonic-hadronic di-$\tau$
decays: $\,\tau_{e} \,\tau_{h}\,$ or $\,\tau_{\mu} \,\tau_{h}\,$.
Then the full accesible final states are: $e\,e$, $\,e \,\mu$,
$\,e \,\tau_{h}$, $\,\mu \,\mu $, $\,\mu \,\tau_{h}$.
Hadronic decays of taus result in jets that must be distinguished
from jets arising from QCD processes. In this case the
``$\tau$-jetiness'' is ensured by the isolation criteria applied
around the second track at Level 3. As a corollary, this prevents
the track from being a product of a light quark or heavy flavored
quark jet.

\noindent Because of the XFT track requirement for both the lepton
and the second track, this trigger is restricted to operate in the
central region of the CDF-II detector. Extending its geometrical
acceptance will represent a probable future upgrade.


\subsection{Central Electron Plus Track Trigger}

The selection of the Electron Plus Track Trigger starts at L1, by
requiring a single EM tower with transverse energy\,\, ($E_{\rm
T}$) \,\,above \,\,\,8~GeV\,\, and\,\, an\,\, associated \,\,XFT


\noindent
\begin{table*}
\begin{center}
\begin{tabular}{|l|l|}
\hline\hline
{\bf Level 1} &  \underline{Electron Object}\\
              &   (EM Shower with XFT track)\\
&\quad $E_{\rm T}(e) > 8$~GeV, $E_{\rm HAD}/E_{\rm EM} < 1/8$ \\
& \quad Associated XFT track $P_{\rm T}^{XFT} \ge 8.34$~GeV$/c$ \\
\hline\hline
{\bf Level 2} &  \underline{Electron Object}\\
& \quad $E_{\rm T}(e) > 8$~GeV,\ $E_{rm XCES}$ $> 2$~GeV\\
& \quad $E_{\rm HAD}/E_{\rm EM} < 1/8$\\
& \underline{2nd XFT track} \\
& \quad XFT track $P_{\rm T}^{XFT}> 5.18$~GeV$/c$\\
& \quad $|\Delta\phi(e,{\rm track})| > 10^{\circ}$\\
\hline\hline
{\bf Level 3} & \underline{Electron Object}\\
              & (electron matched to EM Shower)\\
& \quad $|\Delta z_{\rm CES}| < 8$~cm, $\chi^2_{\rm CES} < 20$\\
& \quad $E_{\rm T} >8$~GeV, $P_{\rm T}>8$~GeV$/c$\\
& \underline{$\tau$-cone track requirements}\\
& \quad $P_{\rm T} \ge 5$~GeV$/c$, $|\eta| \le 1.5$\\
& \quad $N_{\rm track}^{10^{\circ}-30^{\circ}} = 0$\\
& \quad (with ($P_{\rm T}>1.0$~GeV$/c$ and $|\Delta z| < 15$~cm)\\
& \underline{Electron + $\tau$-cone track Object}\\
& \quad $|\Delta z_{0}|\,=\,|z_{0}(e)-z_{0}(trk)| \le 15$~cm,
$|\Delta R| > 0.175$\\
 \hline\hline
\end{tabular}
\label{Table2}\caption{Central Electron Plus Track Trigger.}
\end{center}
\end{table*}

\noindent track with $P_{\rm T}\,>\,\,8$~GeV$/c$. At Level 2 the
electron identification cuts are tightened by requiring the CES
$E_{\rm T}\,>\,\,2$~GeV. In addition, L2 demands a second XFT
track of $5$~GeV$/c$. Level 3 refines these conditions and further
more requires a charged track isolation around the track
reconstructed at L3. Table~2 gives L1, L2 and L3 definitions for
the Electron Plus Track Trigger. The current L3 average
cross-section for the Central Electron Plus Track Trigger is $\sim
29$ nb.

\subsection{Central Muon Plus Track Trigger}

The selection of a muon plus track candidate starts at L1 by
requiring the presence of hits in the CMP chambers associated with
a CMU stub with $P_{\rm T}\,>\,\,6.0$~GeV$/c$ that matches an XFT
track. L2, contrary to the electron case, is at present just an
"auto-accept" as the L2 muon system is not currently operational.
Then L3 requires another track of at least $5$~GeV$/c$ and with an
isolation compatible with the one of a hadronically decaying
$\tau$. The current average cross-section for the Central Muon
Plus Track Trigger is $\sim 16$~nb.

\subsection{CMX-muon Plus Track Trigger}

The Central Muon eXtension (CMX) is a set of drift tubes
sandwiched between two scintillator layers (CSX) realized to give
a further coverage for the muons in the $\eta$ range between $0.6$
and $1$. The CMX Plus Track Trigger then is the a kind of Muon
Plus Track Trigger in the pseudorapidity region:
$0.6\,<\,|\eta|\,<\,1.0$.

\section{Other Tau Triggers}

\subsection{Di-Tau Trigger}

At L1 this trigger requires 2 calorimeter towers with $E_{\rm T}>$
$5$~GeV and 2 matching XFT tracks with $P_{\rm T}>$ $6$~GeV/$c$,
with an angle of $\phi>30^{\circ}$ between them. Level 2 checks
for calorimeter clusters with $E_{\rm T}>$ $10$~GeV and imposes a
track isolation cut. At L3, using the full reconstruction code, 2
$\tau$ candidates with a seed track with $P_{\rm T}>$ $6$~GeV/$c$,
originating from the same vertex: $|\Delta z|<\,10$~cm, are
required. At the present the average cross section for this
trigger is $\sim\,12$~nb.

\subsection{Tau Plus \,\,$\mathbf{\not\!\!\!E_{\rm T}}$ Trigger}

Another Tau Trigger that is not part of the LPT trigger is the Tau
Plus $\not\!\!\!E_{\rm T}$ Trigger. It requires the\,\, presence
of


\noindent
\begin{table*}
\begin{center}
\begin{tabular}{|l|l|}
\hline\hline
{\bf Level 1} &  \underline{Muon Object}\\
              &   (muon stub with XFT track)\\
&\quad $P_{\rm T}(\mu)^{stub} > 6$~GeV$/c$\\
& \quad Associated XFT track $P_{\rm T}^{XFT} \ge 4.09$~GeV$/c$ \\
\hline\hline
{\bf Level 2} & \underline{Muon Object}\\
& \quad Auto Accept\\
& \underline{2nd XFT track} \\
& \quad XFT track $P_{\rm T}^{XFT}>\, 8.0$~GeV$/c$\\
\hline\hline
{\bf Level 3} & \underline{Muon Object}\\
& \quad $|\Delta x_{\rm CMU}| < 15.0$~cm,  $|\Delta x_{\rm CMP}| <
20.0$~cm\\
& \quad $P_{\rm T}>\,8$~GeV$/c$\\
& \underline{$\tau$-cone track requirements}\\
& \quad $P_{\rm T} \ge\,5$~GeV$/c$, $|\eta| \le 1.5$\\
& \quad $N_{track}^{10^{\circ}-30^{\circ}} = 0$\\
& \quad (with $P_{\rm T}>1.5$~GeV$/c$ and $|\Delta z| < 15$~cm)\\
& \underline{Muon + $\tau$-cone track object}\\
& \quad $|\Delta z_{0}|\,=\,|z_{0}(\mu)-z_{0}(trk)| \le 15$~cm,
$|\Delta R| \ge 0.175$\\
 \hline\hline
\end{tabular}
\label{Table2}\caption{Central Muon Plus Track Trigger.}
\end{center}
\end{table*}
\noindent missing transverse energy \,$\not\!\!\!E_{\rm
T}\,>\,10$~GeV at L1. At L2 this request is increased to a value
of \,\,$\not\!\!\!E_{\rm T}>\,20$~GeV and a calorimeter cluster
with an isolated track with $P_{\rm T}\,>\,10$~GeV/$c$ is
required. This is followed by the full event reconstruction at
Level 3, requiring the presence of at least one $\tau$ candidate
having a seed track $P_{\rm T}\,>\, 4.5$~GeV/$c$. At the present,
the Tau Plus\,\, $\not\!\!\!E_{\rm T}$ Trigger average cross
section is $\sim\,5$~nb.

\section{Physics topics addressable by the LPT trigger}

The main goal of this section is to show possible applications of
the LPT trigger~\cite{cdf6325}. Low $P_{\rm T}$ dileptons are a
basic element for many and very important signatures, both in SM
physics and in searches for physics beyond the SM. The main goal
of this section is then to give a mini-review of the possible
applications of the LPT Trigger. As we saw, our trigger is a class
of low momentum dilepton triggers able to access the following
final states: $e\,e$, $\,e \,\mu$, $\,e \,\tau_{h}$, $\,\mu \,\mu
$ and $\,\mu \,\tau_{h}\,$.

\subsection{Standard Model Physics}

\subsubsection{Drell-Yan Production}

The Drell-Yan (DY) process can be accessed for all lepton pairs
produced down to rather low $P_{\rm T}$ and $E_{\rm T}$. It
includes all charged dileptons: $ee$, $\mu\mu$, $\tau\tau$. The
interest to study such production goes beyond the study of the
Drell-Yan process itself as this signature is also typical of
theories beyond the SM such as, in particular, Extra Dimension
Theories.
Also of interest is a study of $Z^{0} \to\tau\tau$ where one of
the $\tau$'s decays leptonically and the other hadronically. In
this respect, this trigger is quite unique to perform this
measurement. This signature is crucial in the search for Higgs(es)
where the Higgs decays into a pair of $\tau$'s as the $Z^{0}
\to\tau\tau$ is a primary source of background events.
The study of the $Z^{0} \to b\bar{b}$ channel is mandatory prior
to the search for the $H^0 \to b\bar{b}$. Also this channel can be
studied with the LPT trigger but only in the case where both the
$b$'s decay semileptonically in $e$, $\mu$ or $\tau$-leptons.

\subsubsection{Top Quark Physics}

Another important physics topic reachable by the LPT trigger is
top quark physics. After the discovery of the top-quark in Run I
of the Tevatron, Run II will provide the unique opportunity to
start doing top-physics both from the point of view of the SM and
to study it, as a background, for many possible new phenomena.
Being able to measure accurately its mass, its cross-section and
its various branching ratios is fundamental. This trigger allows
us to study the production of a top pair through the decay
products into two charged leptons (including the $\tau$-lepton),
thus the various channels: $ee, e\mu, \mu\mu, e\tau, \mu\tau,
\tau\tau$ can be examined. The remark on the isolation constraint
applies again here: the search for the lepton + jet signatures
with  jets other than $\tau$-jets, will be dramatically impeded
with the present trigger.
The LPT trigger could also contribute to the study of the single
top production, especially in the case where the $W$ decay product
of the single top decays into a $\tau$-lepton. Furthermore, the
same trigger gives the possibility to study the $b$-fragmentation
that impacts on the top measurement.
Comparing with the other triggers that are currently used for the
top studies, this trigger gives a unique access to the top
searches where top decays leptonically producing one or two
$\tau$-leptons.

\subsubsection{$W$ and $Z^{0}$ pairs and $H$ production}

For the $W$ or $Z^{0}$ pair production, this trigger selects the
signatures where both $W$ or at least one of the $Z^{0}$ decays
leptonically, including $\tau$'s. It gives a possibility to select
signatures of $WH$ production with Higgs decaying into $b\bar{b}$
or $\tau^+\tau^-$,
and a $W$ through its leptonic decay including $\tau$'s; the other
part of the trigger signature can be used for the Higgs
identification. It also serves to identify $Z^{0}H$ where the
$Z^{0}$ decays leptonically and $H$ decays into a  $b\bar{b}$ or a
$\tau^+\tau^-$. Or one can look for $Z^{0}H$ production with any
decay for the $Z^{0}$ and using the lepton + track signature to
identify the Higgs. The $b$'s can be identified through their
semi-leptonic decays and the $\tau$'s if both decay leptonically,
or if one decays hadronically and the other one leptonically.
The advantage in this case of using the LPT trigger is that it
goes down to rather low $E_{\rm T}$ and $P_{\rm T}$ thresholds
even for processes where the decay products have rather high
$E_{\rm T}$, it thus provides a useful overlap region between the
corresponding standard process backgrounds and the signal to be
looked for. It furthermore permits to study, with lower $E_{\rm
T}$ and $P_{\rm T}$ samples, some systematic effects, such as fake
leptons based on the same trigger selection.

\subsection{Beyond the Standard Model}

The dilepton signature is quite a fundamental signature for many
exotic processes. It is also the basis signature for multilepton
signatures that include at least three charged leptons in the
final state. Here below we briefly describe the main physics
topics addressable from the LPT trigger. For what concerns the
general SUSY scenarios, it should be noticed that the
$\tau$-enriched signature are expected to become dominant as soon
as tan$\beta$ increases. This was one of stronger motivations to
develop such kind of triggers.

\subsubsection{Search for $\mathbf{\tilde{\chi}^{\pm}}$ and  $\mathbf{\tilde{\chi}^{0}}$ associated production}

LEP 200 performed impressive and rather unique work in searching
for the lightest SUSY particle (LSP), namely the neutralino
$\tilde{\chi}_1^0$, in the Minimal Extension of Supersymmetric
Standard Model (MSSM) framework. Tevatron was not able to compete
with LEP results because of the lack of luminosity. One of the
main Run-II challenges is to overcome the LEP limits on gauginos.
CDF at Run I \cite{Ref:Cdf4371} demonstrated that with a powerful
tracking system, the measurement of multilepton signatures with 2
or 3 well measured leptons is an essential tool to search for
chargino-neutralino production: $p\bar{p} \to
\tilde{\chi}_2^0\tilde{\chi}_1^{\pm} \to \textrm{2 or 3 charged
leptons}$. Thus, the LPT trigger allows to select this class of
events including $\tau$'s in the final states.

\subsubsection{Search $\mathbf{\tilde{g}}$ and $\mathbf{\tilde{q}}$ cascade decays}

The higher Run II integrated luminosity together with the improved
CDF-II detector performances will allow the search for gluino and
squark  cascade decays, otherwise non searchable as it was during
the Run I.
Among the interesting possibilities, there is the case where the
gluino decays into a $b$-quark and sbottom ($\tilde{b}$), followed
by $\tilde{b}\to b\tilde{\chi}_2^0$ ($\tilde{\chi}_2^0 \to
\tau\tilde{\tau}$ and $\tilde{\tau}\to\tau\tilde{\chi}_1^0$).
This $\tau$-enriched signature will become more dominant as
tan$\beta$ becomes larger. The final decay product of the gluino
will be: $bb\tau\tau\tilde{\chi}_1^0$. The $\tau$'s produced in
this case will have relatively low $P_{\rm T}^{\tau}$, which makes
the LPT trigger particularly sensitive to this channel.

\subsubsection{Search for scalar top and $\mathbf{\mathcal{R_{P}}}$ Violation}

The search for the scalar partner of the top quark\,
$\tilde{t}_{1}$\, in the MSSM framework is generally performed
looking at the\, $c\,\tilde{\chi}^{0}$\, or\,
$b\,\tilde{\chi^{\pm}}$ final state, that means essentially trying
to tag the $b$-jet or the $c$-jet.
In models with  $\mathcal{R_{P}}$ Violation (RPV) the stop squark
can decay to $b\tau$ giving the following final states: \,$p
\bar{p} \,\rightarrow \,\tilde{t}_{1}\, \bar{\tilde{t}}_{1}
\,\rightarrow\, \tau_{h} \,b \,\tau_{\ell}\, \bar{b}$\, that proof
that LPT trigger is instrumental in this case.
As studied in Run 1 and LEP 200, various $\mathcal{R_{P}}$
violated SUSY scenarios lead to enriched multilepton signatures.
For all these cases, this trigger will be quite useful. In
addition, the fake leptons that are a crucial issue for these
searches can be studied with the same triggered sample of data,
and thus reduce systematic effects.

\subsubsection{Search for non-SM top quark decays}

One of the major issues of Tevatron will be to understand the real
nature of top quark production and decay. Top quark production, be
it $t\bar{t}$ or single top, is an ideal place where to look for
new physics.  If there is any new physics associated with the
generation of mass, it may be more apparent in the top quark
sector than with any of the other lighter, known, fermions.  Many
models predict new particles or interactions that couple
preferentially to the third generation and in particular to the
top quark.  These models extend the strong, hypercharge or weak
interactions in such a way that, at some scale, the new groups
spontaneously break into their SM subgroup: $SU(3)_h\times
SU(3)_l\to SU(3)_C$, $SU(2)_h\times SU(2)_l\to SU(2)_{W}$, and
$U(1)_h\times U(1)_l\to U(1)_Y$, where $h$ represents the third
(heavy) generation and $l$ the first two (light) generations.  As
a result, one would expect production rate and kinematic
distributions of the decay products to differ from the SM
predictions. The SM predicts ${\mathcal BR}(t \rightarrow
bW)\,>\,\,0.998$. Other decays allowed in the SM are not only
rare, but also too difficult to disentangle from backgrounds to be
observed in the future. Nevertheless, one must try to be sensitive
to all conceivable signatures of top quark decay, as some can be
enhanced by several orders of magnitude in scenarios beyond the
SM. Here we highlight some scenarios, with interesting theoretical
motivations, in the rich of Tevatron Run 2, that will get a
relevant advantage of the LPT trigger.


\subsubsection{Search for SUSY decays of top}

The CDF experiment has already searched, during the Run I, for
events of this type where the SM top decay proceeds as $t\,\to\,
W\,b\, \to\,$ $\ell\,\nu_\ell \,b $ $\;$ ($\ell=\,e,\,\mu$), while
the SUSY decay of the other top proceeds as $t \;\to\; \tilde{t}_1
\,\tilde\chi_1^0 \;\to\; b \,\tilde \chi_1^+\,\tilde\chi^0_1\;
\to\; b\, q_1\, \bar q_2 \, \tilde\chi^0_1 \, \tilde\chi^0_1$
setting a 95\% C.L. limit on ${\mathcal BR}(t \to \tilde{t}_1
\tilde\chi_1^0)$ as function of $\,m_{\tilde t_1}\,$,
$\,m_{\tilde\chi_1^\pm}$ and
$\,m_{\tilde\chi^0_1}\,$~\cite{top_stop}. The LPT trigger will
allow us to access also final states containing $\tau$'s, making
possible to increase the sensitivity on the branching ratio
${\mathcal BR}(t \to \tilde{t}_1 \tilde\chi_1^0)$.

\subsubsection{Top decays in BRPV}

Another class of models that will be possible to investigate by
using the LPT trigger are the Bilinear $\mathcal{R_{P}}$ Violation
(BRPV)~\cite{Ref:NuPhysB524, Ref:hepph9711435, Ref:NuPhysB451,
Ref:PhysRevD51}. These models are well-motivated theoretically as
they arise as effective truncations of models where $R$--Parity is
broken spontaneously \cite{Ref:PhysLetB251} through right handed
sneutrino vacuum expectation values (vev) $\tilde{\nu}^c = v_R
\neq 0$.
In the BRPV models the charginos mix with the charged leptons, the
neutralinos with neutrinos, and the charged sleptons with the
charged Higgs boson \cite{Ref:NuPhysB524, Ref:hepph9711435,
Ref:NuPhysB451}. Therefore, the top can have additional decay
modes:

\[
t \to {\tilde \tau}^+_1 \, b \, , \hspace{4mm} t \to \nu_\tau \,
{\tilde t}_1 \, , \hspace{4mm} t \to \tau^+ \, {\tilde b}_1
\nonumber
\]

\noindent In every case the various decay modes lead to cascade
decays:

\begin{center}
\begin{tabular}{lll}
$t \to {\tilde \tau}^+_1 \, b$  & $\to \tau^+ \, \nu_\tau \, b$
& \\
                    & $\to \tau^+ \, {\tilde \chi}^0_1 \, b$    & $\to \tau^+   \, f \, \bar{f}  \, \nu_\tau \, b$ \\
                &                       & $\to \tau^+   \, f \, \bar{f}' \, \tau^\pm \, b$ \\
                    & $\to \nu_\tau \, {\tilde \chi}^+_1 \, b$  & $\to \nu_\tau \, f \, \bar{f}' \, \nu_\tau \, b$ \\
                    &                       & $\to \nu_\tau \, f \, \bar{f}  \, \tau^+ \, b$ \\
                    &  $\to c \, s \, b$                & \\
$t \to \tau^+ \, {\tilde b}_1$  & $\to \tau^+ \, \nu_\tau \, b$
& \\
                    & $\to \tau^+ \, {\tilde \chi}^0_1 \, b$    & $\to \tau^+ \, f \, \bar{f}  \, \nu_\tau \, b$ \\
                    &                       & $\to \tau^+ \, f \, \bar{f}' \, \tau^\pm \, b$
\end{tabular}
\end{center}

\vskip .2cm \noindent In nearly all cases there are two $\tau$'s
and two $b$-quarks in the final state plus the possibility of
additional leptons and/or jets. Therefore, $b$-tagging and a good
$\tau$ identification are important for extracting these final
states. The background will come mainly from the production of one
or two gauge bosons plus additional jets.

\subsubsection{Search for SUSY Higgs}

The search for the lightest SUSY Higgs ($h_0$) will be very
similar to the case already mentioned of the SM Higgs search. The
difference can be in an enhanced rate in particular for the case
where the Higgs decays into a $\tau$-pair. This trigger is quite
unique for detecting $h^0 \to \tau^+\tau^-$ with one leptonic
$\tau$ and one hadronic $\tau$ or both leptonic.

\subsubsection{Search for Lepton Flavor Violating Higgs decays}

Strong evidence in favor of neutrino masses and mixing, obtained
in Super-Kamiokande and other neutrino experiments, opened a new
epoch. Several extensions of the SM, Supersymmetric and not,
assume the existence of flavor-changing couplings of a Higgs
boson. Among them, the generic Two Higgs Doublet Model (THDM-III)
can be taken as a representative case where  ${\mathcal BR}(H\to
\tau \mu )$ can reach values of order $\simeq 10^{-1}-10^{-2}$ as
shown in~\cite{Ref:C010630}.
The search for Lepton Flavor Violating (LFV) Higgs decay, $H\to
\tau \mu$, is not only well motivated by the favorable
interpretation of the $\nu_{\mu}-\nu_{\tau}$ oscillation but also
has the unique advantage to be at the same time a reachable Higgs
discovery channel and a way to constrain the present loose bounds
on the size of the LFV factor $\kappa_{\tau\mu}$ by following a
totally model independent search for LFV Higgs decays. Naturally
this is an interesting signature for which the use of the Lepton
Plus Track Trigger is mandatory.

\subsubsection{Search for Large Extra Dimensions}

Few years ago, Arkani-Hamed, Dimopoulos and Dvali (ADD) suggested
that the foundamental quantum gravity scale is of the order of the
Fermi scale: $G_{\mathcal{N}}= G_{\mathcal{D}}/R^{N}$ where
$G_{\mathcal{D}}$ is the microscopic Newton constant, N is the
number of extra spatial dimensions, hiden, as compactified at each
point of the $4-$dimensional space, with a compactification radius
R~\cite{HDD1}. Present gravity experiments, with Cavendish-type
setups, cannot test gravity below the mm scale. Tevatron can study
gravity below this edge. As a matter of fact, the virtual exchange
of graviton towers ( $G_{\rm KK}$) either leads to modifications
in SM cross sections and asymmetries or to new processes not
allowed in the SM at the tree level. In the case of virtual
$G_{\rm KK}$ emission, gravitons  lead to apparent violation of
$4$-momentum as well as of the angular momentum. The impact of
virtual gravitons then can be observed in processes such as: $q
\bar{q} \rightarrow G_{\rm KK} \rightarrow \gamma \gamma$ or $g g
\rightarrow G_{\rm KK} \rightarrow \ell^{+}\ell^{-}$ where the ADD
model introduces production mechanism that can increase the
cross-section~\cite{HDD2}. The characteristic signatures of
virtual graviton exchange correspond then to the formation of
massive systems abnormally beyond the SM expectations. The LPT
trigger gives a very relevant access to this physics as it permits
to study pairs of charged leptons including $\tau$'s.


\section{Preliminary results: the $\mathbf{ Z^{0} \rightarrow \tau \tau}$ signal}

As a prelude to any future analysis, and in order to illustrate
the technique, we discuss, in this section, the first look at
$Z^{0} \rightarrow \tau \tau$ events collected with the Electron
Plus Track Trigger. The data used for this preliminary analysis
was collected between March 2002 and January 2003, corresponding
to a total integrated luminosity of $\sim$ $72 \;{\rm
pb}^{-1}$~\cite{cdf6402}. In this case we will have one tau
decaying leptonically in an electron ($\tau_{e}$) and the other
one hadronically ($\tau_{h}$). Our analysis begins by requiring
the presence of at least one good electron in the central region
($|\eta^{e}|< 1.0$) with $E_{\rm T}\,\ge\,10$~GeV and $P_{\rm T}\,
\ge\, 8$~GeV$/c$. The cuts applied for the electron identification
are similar to the standard CDF-II cuts for selecting inclusive
high $E_{\rm T}$ central electrons.
Since the second leg in the decay $Z^{0} \,\rightarrow\, e\, e$
can be misidentified as one prong $\tau_{h}$, we strictly remove
any possible Drell-Yan $Z^{0} \rightarrow e^{+} e^{-}$ candidate
event by using both calorimeter and track information.
Then, we require the event to have at least one $\tau_{h}$ object
in the central region $|\eta^{\tau_{h}}|\,<\,1.0$ with a cluster
transverse energy: $E^{cluster}_{\rm T}(\tau_{h})\,>\,20$~GeV. The
SM backgrounds for  \,$Z^{0} \rightarrow \tau_{e} \tau_{h}$\,
analysis are essentially
\noindent
\hspace{-0.3cm}\epsfxsize=\columnwidth\epsffile{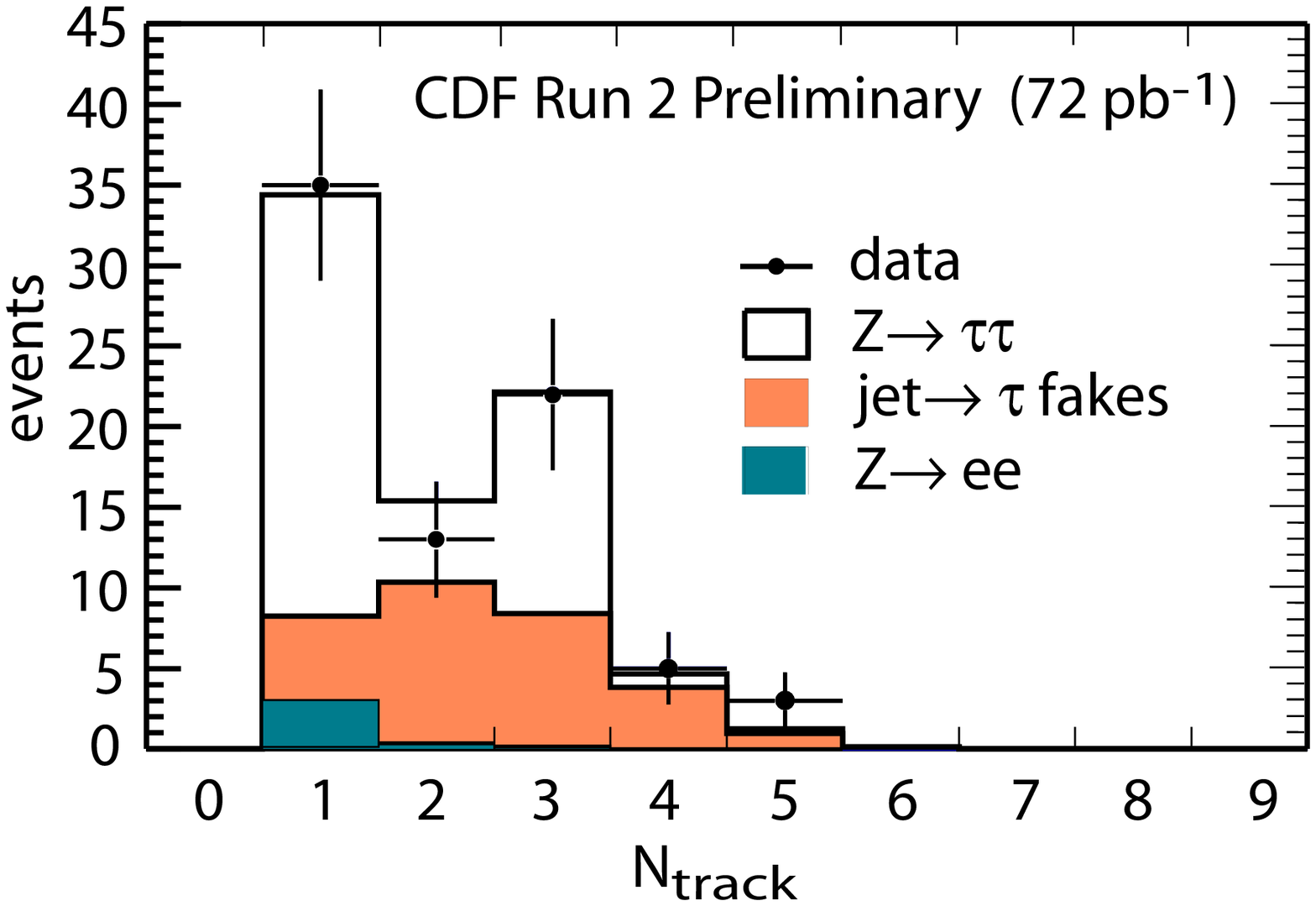}
Figure 4: {\it Track multiplicity of $\tau$ objects in the
Electron Plus Track Trigger dataset after baseline selections and
$M_{\rm T}(e,\not\!\!\!E_{\rm T})$ and $P_{\rm
T}(e,\not\!\!\!E_{\rm T})$ cuts. Data are shown together with the
$Z^{0} \rightarrow \tau^{+} \tau^{-}$, $W\,+\,$jets\, and $Z^{0}
\rightarrow e^{+}e^{-}$\, expected contributions.}
\noindent{\footnotesize }%
\vskip15pt
\noindent the\,\, QCD\,\,\,\, $W(\rightarrow e\nu)\,\,+\,\,$jets,
\,$Z^{0}/\gamma^{*}\,\rightarrow\,e^{+}\,e^{-}$, \,\,\,\,$W\,W$
 diboson production and \,$t \bar{t}$ events.
The QCD \,$W\,+\,$jets \,background is partially reduced by
applying a cut on the transverse mass: $M_{\rm T}(e,
\not\!\!\!E_{\rm T})\le 25$~GeV$/c^{2}$. In fact, the $W$ events
show up in the higher mass region as in this case $M_{\rm T}(e,
\not\!\!\!E_{\rm T})$ is the transverse mass of the $W$.
After the $M_{\rm T}(e, \not\!\!\!E_{\rm T})$ cut, a significant
number of QCD background events is still present in the sample. A
further cut is then applied $P_{\rm T}(e, \not\!\!\!E_{\rm
T})\,\ge\,25$~GeV$/c$.
After applying the baseline selection cuts for the electron and
for the $\tau_{h}$ candidate and the further cuts on $M_{\rm T}(e,
\not\!\!\!E_{\rm T})$ $P_{\rm T}(e, \not\!\!\!E_{\rm T})$ we end
up with a sample containing 78 events. This sample is expected to
consist of $\sim 59$ \% of signal events with the rest of the
background events dominated by QCD jet production.
The track multiplicity associated with the $\tau_{h}$ candidate
found in events passing selection cuts is shown in Figure~4. We
observe a clear $\tau$ signal (1,3 prong signature) over
background levels even before requiring the opposite sign charge
for the electron and the $\tau_h$.
If we define the charge of the $\tau_{h}$ candidate as the sum of
the track charges: $Q(\tau_{h})$ $
=\,\sum_{j=1}^{N^{trk}}q^{trk}_{j}$ it is possible to classify the
sample of 78 events in same sign


\noindent (SS) and opposite sign (OS) events. We are left with 47
OS events. Figure~5 shows the mass of the system containing an
electron, a hadronically decaying $\tau$ and the $\not\!\!\!E_{\rm
T})$. As it is possible to see, the mass distribution of the
electron-$\tau_h$- $\not\!\!\!E_{\rm T}$ system
($M(e\,+\,\tau_{h}\,+\,\not\!\!\!E_{\rm T})$)
 in the data is consistent with the $Z^{0} \rightarrow \tau\tau$\,
hypothesis. To study and collect a sample of $Z^{0} \rightarrow
\tau \tau$ is an important benchmark for several reason. This
sample will be fundamental in

\noindent
\vskip15pt\hspace{-0.8cm}\epsfxsize=\columnwidth\epsffile{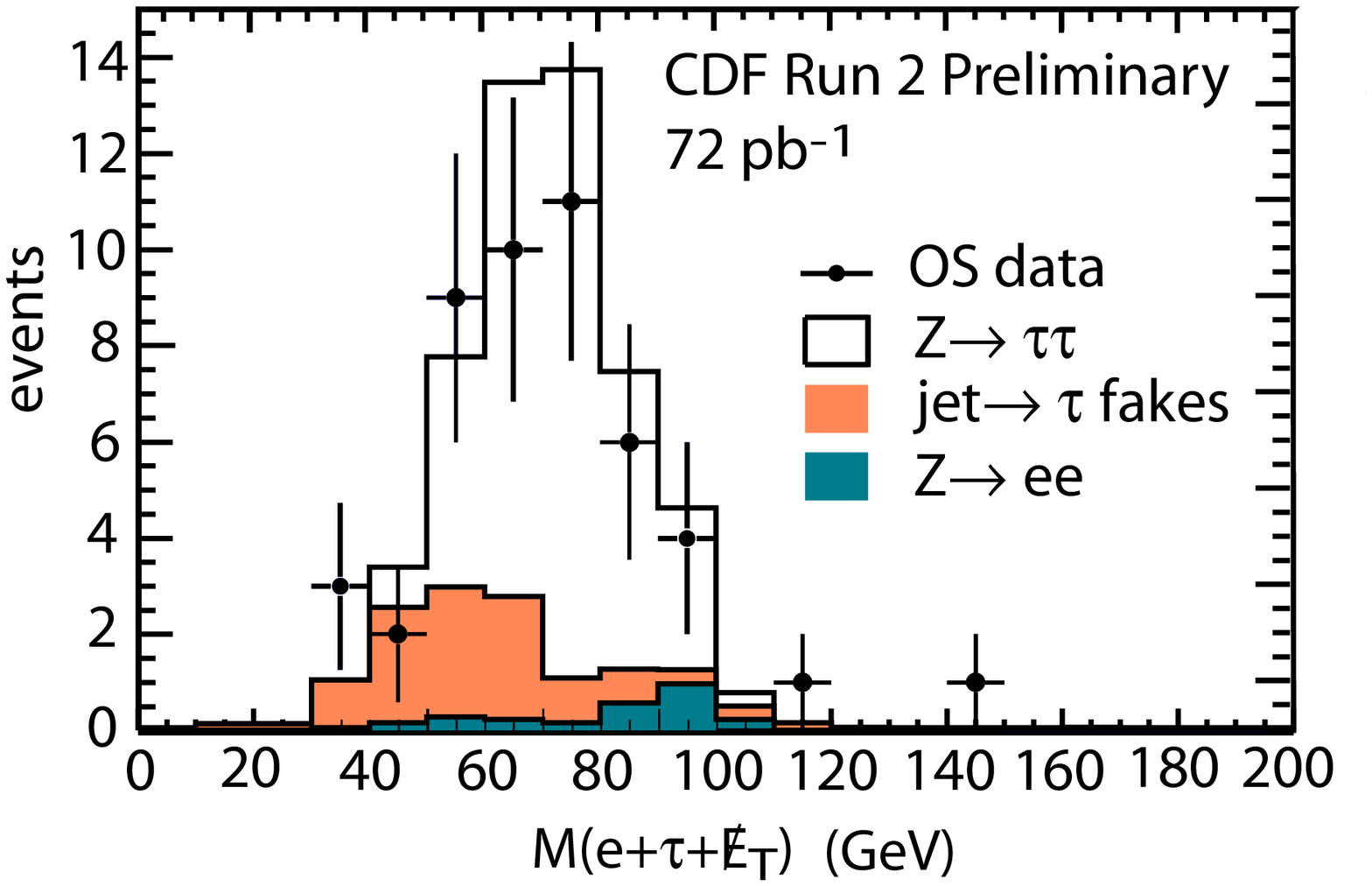}
\vskip12pt\noindent Figure 5: {\it Mass distribution of the
opposite sign $e$, $\tau_{h}$ events in the Electron Plus Track
Trigger dataset after baseline, $M_{\rm T}(e,\not\!\!\!E_{\rm T})$
and $P_{\rm T}(e,\not\!\!\!E_{\rm T})$ cuts. Data are shown
together with the $Z^{0} \rightarrow \tau^{+} \tau^{-}$,
$W\,+\,$jets\, and $Z^{0} \rightarrow e^{+}e^{-}$\, expected
contributions.}
\noindent{\footnotesize }%
\vskip15pt
\noindent order to calibrate the trigger system and to understand
the system global performances. On the other hand this preliminary
analysis is important as the $Z^{0} \rightarrow \tau \tau$ events
are one of the most significant backgrounds for many of the
analysis described in the previous section.

\section{Summary}

Without any doubt $\tau$ physics will be one of the most
intriguing chapters in the Run II physics program at Tevatron.
Hence, the possibility to select tau enriched samples already at
trigger level open new perspectives for CDF-II experiment. In this
paper we described basic ideas, implementation and performances of
dedicated triggers, allowing to select events with at least one $
\tau$ candidate in the final state.
With the first results on $Z^{0} \rightarrow \tau \tau$
production, obtained with one of the described trigger, we start
to investigate a wide and intense area of physics searches with
$\tau$ leptons in the final state.

\section{Acknowledgments}

We thank the Fermilab staff and the technical staff of the
participating Institutions for their contributions. The work
described in this paper was supported by the U.S. Department of
Energy and National Science Foundation; the Istituto Nazionale di
Fisica Nucleare; the Ministry of Education, Culture, Sports,
Science and Technology of Japan, the National Science Council of
the Republic of China; the Swiss National Science Foundation; the
A.P. Sloan Foundation; the Bundensministerium fur Bildung un
Forshung; the Korea Science and Engineering Foundation (KoSEF);
the Korea Reasearch Foundation; and the Comision Interministerial
de Ciencia y Tecnologia, Spain.
\noindent We thank Laszlo Jenkovszky and his stuff for the kind
and enjoable hospitality.


\end{multicols}
\end{document}